\documentclass[12pt,a4paper,showkeys]{article}
\usepackage[font=footnotesize,labelfont=bf,width=1.1\textwidth]{caption}
\usepackage{authblk}
\usepackage{epsfig,amsmath,amsfonts,amssymb,cancel}
\usepackage{graphicx,latexsym,bbold,mathbbol,color}
\usepackage{stackrel,epstopdf,xcolor,hyperref,cite}
\usepackage{slashed}
\usepackage[hang,flushmargin]{footmisc}
\usepackage{soul}
\allowdisplaybreaks

\newcommand{\Bpsi}{\bar \Psi}
\newcommand{\bq}{\bar Q}
\newcommand{\ssl}{\slashed}

\newcommand{\bp}{p_{_{\!-}}}

\begin{document}
\numberwithin{equation}{section}

\title{Dressed Dirac Propagator from a Locally Supersymmetric ${\cal N}=1$ Spinning Particle}

\author[a,b,c]{Olindo Corradini\,\footnote{E-mail: olindo.corradini@unimore.it}}
\author[a,d]{Gianluca Degli Esposti\,\footnote{E-mail: 216395@studenti.unimore.it}}

\affil[a]{{\small\it Dipartimento di Scienze Fisiche, Informatiche e Matematiche, \protect\\[-1.5mm]
Universit\`a degli Studi di Modena e Reggio Emilia, 
Via Campi 213/A, I-41125 Modena, Italy}}

\affil[b] {{\small\it INFN, Sezione di Bologna,  Via Irnerio 46, I-40126 Bologna, Italy}}

\affil[c]{{\small\it Max-Planck-Institut f\"ur Gravitationphysik, Albert-Einstein-Institut,\protect\\[-1.5mm] Am M\"uhlenberg 1, 14476 Golm, Germany}}

\affil[d]{{\small\it Helmholtz-Zentrum Dresden-Rossendorf, \protect\\[-1.5mm] Bautzner Landstra\ss e 400, 01328 Dresden, Germany}}

\date{\empty}

\maketitle

\thanks{\begin{center}\small This article is registered under preprint number, arXiv:2008.03114 [hep-th]\\[1cm] \end{center}}

\abstract{We study the Dirac propagator dressed by an arbitrary number $N$ of photons by means of a worldline approach, which makes use of a supersymmetric ${\cal N}=1$ spinning particle model on the line, coupled to an external Abelian vector field. We obtain a compact off-shell master formula for the tree level scattering amplitudes associated to the dressed Dirac propagator. In particular, unlike in other approaches, we express the particle fermionic degrees of freedom using a coherent state basis, and consider the gauging of the supersymmetry, which ultimately amounts to integrating over a worldline gravitino modulus, other than the usual worldline einbein modulus which corresponds to the Schwinger time integral. The path integral over the gravitino reproduces the numerator of the dressed Dirac propagator. \\[1cm]}

\noindent

\newpage


\section{Introduction}
\label{sec:intro}
The use of first-quantized methods to compute scattering amplitudes and effective actions has been known since the renowned work of Feynman~\cite{Feynman, Feynman2}, who proposed a worldline representation for the scalar QED propagator, which even included one-loop photon self energy contributions.  

However, it was only in the early nineties that these methods were taken seriously for Quantum Field Theory (QFT) computations, as an alternative to second-quantized approaches, namely after the work of Bern and Kosower~\cite{Bern}, who derived a compact master formula for one-loop gluon scattering amplitudes, from the infinite tension limit of string theory, and after Strassler's work~\cite{Strassler}, who rederived Bern-Kosower master formula for QED amplitudes, directly from particle path integrals. After these seminal contributions there have been numerous applications of the method---see the papers~\cite{Schubert, Edwards-Schubert} for a self-contained summary of the worldline applications to quantum field theory computations and for an extensive bibliography. In particular, for QED related physics, these first-quantized ``string-inspired'' methods have found a variety of applications, including multiloop calculations~\cite{Schmidt:1994zj, Schmidt:1994aq}, Euler-Heisenberg lagrangians~\cite{Dunne:2002qf, Dunne:2002qg}, derivative-expansions of the effective action~\cite{Cangemi:1994by, Gusynin:1995bc, Gusynin:1998bt, Ahmadiniaz:2013wva}, as well as the inclusion of constant external electromagnetic fields~\cite{Shaisultanov:1995tm, Adler:1996cja, Reuter:1996zm} and finite temperature computations~\cite{McKeon:1992if, McKeon:1993sh, McKeon:1997pm, Shovkovy:1998xw, Sato:1998hv, Venugopalan:2001hp}, just to name a few.   
Although a generalization of the Bern-Kosower master formula to the open line case for a scalar field was proposed soon after Strassler's worldline papers by Daikouji et al~\cite{Daikouji}, the amount of scientific results concerning the use of worldline techniques to the computation of QFT results which involve open lines, so far, has been much less than that concerning QFT loops. Some exceptions include computations of thermal Green's functions~\cite{McKeon:1993np, McKeon:1993bc, McKeon:1994hd}, open line spin factors~\cite{Karanikas:1999hz}, particle path integral representations for the scalar and Dirac propagators~\cite{Casalbuoni:1974pj, Casalbuoni:1980}, the worldgraph approach~\cite{Dai:2006vj, Holzler:2007xt, Dai:2008bh}, spinning particle representations for Einstein gravity~\cite{Bonezzi:2018box}, and Compton-like scattering for scalar particles coupled to gravity~\cite{Ahmadiniaz:2019ppj}. An important reason why string-inspired methods have developed more for loop computations, is that one-loop effective actions, and their associated one-loop one-particle irreducible correlations functions, are represented in terms of traces of differential operators, which find a rather natural representation in terms of particle models, as already shown by Alvarez-Gaum\'e and Witten in their renowned work about gravitational anomalies~\cite{Alvarez} (see the book~\cite{Bastianelli-PvN} for a recent detailed review of the subject.) In fact, in such cases, spinorial degrees of freedom can often be conveniently represented in terms of a matrix-valued potential inside purely bosonic particle models.     

However, recently, an efficient first-quantized formalism to the computation of a Dirac propagator dressed by external electromagnetism, has been developed~\cite{fppaper1, fppaper2}, which makes use of an ${\cal N}=1$ supersymmetric spinning particle model. In that approach the denominator of the Dirac propagator---referred to as `the kernel'---is represented in terms of a worldline path integral and the fermionic degrees of freedom are obtained by using the spinning particle method, proposed years ago by  Gitman and Fradkin~\cite{Gitman}, who developed earlier  constructions~\cite{Berezin:1975md, Casalbuoni:1975bj, Brink:1976sz}---the specific feature of such formalism is the use of `Weyl symbols'~\cite{Berezin:1975md, Henneaux:1992ig}. In the Refs.~\cite{fppaper1, fppaper2} the numerator---and the final expression for the propagator---is obtained by acting on the denominator with the dressed Dirac operator.  In the present manuscript instead, we take a complementary path and decide to represent the fermionic degrees of freedom of the Dirac particle in terms of a coherent state basis. Moreover, in our approach we gauge the worldline supersymmetry, and the resulting worldline gravitino gauge field  leaves a Grassmann modulus, in the path integral, which inserts the numerator of the (dressed) Dirac propagator. This allows us to obtain a Bern-Kosower-like master formula for the Dirac propagator in momentum space, dressed with the insertion of $N$ photon, which we will refer to as the ``$N$-photon Dirac propagator'' and which, similarly to other worldline master formulas, is valid off the mass-shell of the particles involved.  Another advantage of our derivation, compared to the more conventional Feynman rules based on the space-time QED lagrangian, is that the {\em full} final expression is expressed in terms of products of coherent states eigenvalues, rather than products of gamma matrices, similarly to what happens in the aforementioned method based on the symb map.  The price to pay is the presence of the aforementioned gravitino modulus, which mixes with the open-line boundary conditions, in order to ultimately reproduce the numerator of the dressed propagator. Other auxiliary Grassmann numbers are introduced, which help imposing the multilinearization in the photon polarizations; they can be seen to naturally emerge by describing the spinning particle in terms of a super worldline time (see e.g.~\cite{fppaper1}).        

\section{The Spinning Particle Model}
\label{sec:spinning-particle}
Let us describe the particle model we use to compute our main master formula, which consists of a locally symmetric ${\cal N}=1$ supersymmetric massive spinning particle model, coupled to external electromagnetism. The dynamical variables are phase-space geometric coordinates, $(x^\mu, p_\mu)$, and a set of Majorana coordinates, $\psi^\mu$, plus an additional Majorana variable, $\psi^5$. Although the spinning particle model below can be defined in any
spacetime dimension, in odd spacetime dimensions there is no obvious definition of
chirality. Thus, the worldline representation of massive Dirac
particles in terms of the present wordline spinning particle models
should be restricted to even-dimensional spacetimes. Here,
for definiteness, we stick to four-dimensional spacetime, in which the
Minkowskian phase space action of the model reads,~\footnote{Note that such action can be easily obtained from dimensionally reducing the free five-dimensional spinning particle and covariantizing the momentum.}
\begin{align}
    S[e,\chi;x,p, \psi, \psi^5;A] =\int_0^1 d\tau \Big[p_\mu\dot x^\mu +\tfrac{i}{2}\psi_M\dot\psi^M & -e\underbrace{\tfrac12 (\Pi_\mu \Pi^\mu(A)+m^2+i{\rm e}F_{\mu\nu}\psi^\mu \psi^\nu)}_{H}\nonumber\\
    &-i\chi\underbrace{(\Pi_\mu\psi^\mu +m\psi^5)}_{Q}\Big]
\end{align}
where $\mu=0,\dots, 3$, whereas $M=0,\dots,3,5$, and $\Pi_\mu=p_\mu-{\rm e} A_\mu$ is the covariant momentum. The first class constraints $H$ and $Q$ satisfy the classical ${\cal N}=1$ supersymmetry algebra $\{Q,Q\}_{D.b.}=-2iH$. The Euclidean configuration space action, which we obtain by performing the Wick rotation $ie\to e$ and $i\chi\to\chi$, and by integrating out the particle momentum, reads
\begin{equation}
    S[e,\chi;x, \psi, \psi^5;A] =\int_0^1 d\tau \Big[\tfrac{1}{2e}\big(\dot x^\mu-\chi\psi^\mu\big)^2 +\tfrac{1}{2}\psi_M\dot\psi^M-i{\rm e}\dot x^\mu A_\mu +\tfrac12 e m^2+i\chi m \psi^5+\tfrac{ie}{2}{\rm e} \psi^\mu F_{\mu\nu} \psi^\nu  \Big]~.
    \label{eq:N=1-euclide}
\end{equation}
At the quantum level, in order to build the fermionic Hilbert space, we find it convenient to use the so-called {\it doubling trick}, which consists in doubling the number of Grassmann-odd coordinates, by adding a new set of free variables $\psi'^M$, which satisfy the same Dirac brackets as the $\psi$'s. This thus allows us to complexify the fermion coordinates, $\Psi^M:=\tfrac{1}{\sqrt 2} (\psi^M+i\psi'^M)$ and use the coherent state basis, for the corresponding fermionic operators, i.e. $\langle \bar\lambda| \hat{\bar\Psi}^M = \langle \bar\lambda| \bar\lambda^M,\ \hat \Psi^M |\eta\rangle = \eta^M |\eta\rangle$. However, the coupling in the action will only involve the original Majorana fields, which now read $\psi^M =\tfrac{1}{\sqrt 2} (\Psi^M+\bar\Psi^M)$. 

The dressed Dirac propagator, in the coherent state basis, is thus linked to the line path integral of the model~\eqref{eq:N=1-euclide}, extended with the aforementioned doubling trick, and equipped with the suitable boundary term, i.e.,
\begin{eqnarray}
    S^{x',x}(\bar\lambda,\eta ; A_\mu)&:=&\frac{1}{\langle\bar\lambda|\eta\rangle} \langle x',\bar\lambda |\,
    \frac{\hat{\slashed{\Pi}}(A)+m\hat \gamma^5}{\hat \Pi^2(A)+m^2}\, |x,\eta\rangle\nonumber\\
    &\sim& e^{-\bar \lambda \eta} \int \frac{De D\chi}{\rm Vol.\ Gauge} \int_{x(0)=x}^{x(1)=x'} Dx \int_{\Psi(0)=\eta}^{\bar\Psi(1)=\bar \lambda} D\bar\Psi D\Psi~ e^{-S_{qu}[e,\chi;x, \Psi;A]}~,
\end{eqnarray}
where the quantum action reads
\begin{align}
  S_{qu}[e,\chi;x, \Psi;A] &= \int_0^1 d\tau \Big[\tfrac{1}{2e}\big(\dot x^\mu-\chi\psi^\mu\big)^2 +\bar \Psi_M\dot\Psi^M-i{\rm e}\dot x^\mu A_\mu +\tfrac12 e m^2+i\chi m \psi^5+\tfrac{ie}{2}{\rm e} \psi^\mu F_{\mu\nu} \psi^\nu  \Big]  \nonumber\\
  &-\bar\Psi_M \Psi^M(1)~.
\end{align}
Above, we already divide out the coherent state normalization $\langle\bar\lambda|\eta\rangle =e^{\bar \lambda \eta}$ and the first path integral, over the gauge fields $e$ and $\chi$, is intended to be performed by suitable gauge-fixing and by removing the overall gauge group volume. The gauge transformations for $e$ and $\chi$ in~\eqref{eq:N=1-euclide} reads
\begin{align}
    &\delta e = \dot \xi +i2\chi \epsilon\\
    &\delta \chi = \dot \epsilon
\end{align}
where, $\xi$ and $\epsilon$ are the gauge parameters for the local worldline translation and local worldline supersymmetry, respectively, which on the open line, vanish at the end points $\tau =0,1$. Thus, one can easily show that $e$ and $\chi$ can be gauged away, up to two constant moduli, which are orthogonal to the gauge transformations and are supported by the open-line boundary conditions (on the closed circle, due to antiperiodicity, there is no modulus for the gravitino); namely $e\to 2T$ and $\chi \to \theta$, with $T$ being a positive real number, and $\theta$ a Grassmann number. Hence, we have
\begin{align}
 S^{x',x}(\bar\lambda, \eta ; A_\mu)&:=
\sqrt2 e^{-\bar \lambda \eta} \int_0^\infty dT \int d\theta \int_{x(0)=x}^{x(1)=x'} Dx \int_{\Psi(0)=\eta}^{\bar\Psi(1)=\bar \lambda} D\bar\Psi D\Psi~ e^{-S_{qu}[2T,\theta;x, \Psi;A]}   
\label{eq:path-integral}
\end{align}
and the factor $\sqrt{2}$, is a normalization factor that takes into account that the $\hat \psi^M$'s satisfy the conventional Clifford algebra, up to a factor of 2, i.e. $\hat \psi^M =\tfrac{1}{\sqrt 2} \hat \gamma^M$ and $\hat \gamma^M = \hat{\bar\Psi}^M +\hat{\Psi}^M$~\cite{Casalbuoni:1980}. 

In order to explicitly perform the path integral~\eqref{eq:path-integral}, we find it convenient to use the background field method and split the paths as (indices are raised/lowered with the flat Minkowski metric)
\begin{align}
    x^\mu(\tau)&=x^\mu+(x'-x)^\mu\tau+q^\mu(\tau) \nonumber \\
    \Psi^M(\tau)&=\eta^M+Q^M(\tau) \\
    \Bpsi_M(\tau)&=\bar \lambda_M+\bq_M(\tau) \nonumber
\end{align}
where $q, Q, \bq$ can be interpreted as quantum fluctuations from the classical solutions to the free equations of motion. These fields satisfy the boundary conditions $q^\mu(0)=q^\mu(1)=Q^M(0)=\bq_M(1)=0$. The splitting for Majorana fields reads
\begin{align}
    \psi^M(\tau) = \Upsilon^M +\frac{1}{\sqrt2} (Q^M(\tau) +\bar Q^M(\tau))
\end{align}
where $\Upsilon^M:=\tfrac{\eta^M+\bar \lambda^M}{\sqrt2}$. Thus, the path integral finally becomes
\begin{align}
 S^{x',x}(\bar\lambda, \eta ; A_\mu)&=\sqrt 2
\int_0^\infty dT \int d\theta\, e^{-m^2 T-\tfrac{(x-x')^2}{4T}+\tfrac{1}{2T}\theta \Upsilon_\mu (x'-x)^\mu-im \theta \Upsilon^5}\nonumber\\& \int_{q(0)=0}^{q(1)=0} Dq \, e^{-\tfrac{1}{4T}\int \dot q^2}\int_{Q(0)=0}^{\bar Q(1)=0} D\bar Q DQ\, e^{-\int \bar Q \dot Q} e^{-S'[q,Q,\bar Q; A]}   
\label{eq:path-integral-fin}
\end{align}
where
\begin{align}
  S'[q,Q,\bar Q; A] =\int_0^1d\tau \Big[& -\tfrac{\theta}{2\sqrt 2 T} ((x'-x) +\dot q)^\mu (Q_\mu+\bar Q_\mu)-i{\rm e} (x'-x+\dot q)^\mu A_\mu (x+(x'-x)\tau +q)\nonumber\\
  &+iT {\rm e} \big(\Upsilon^\mu +\tfrac{1}{\sqrt 2}(Q^\mu +\bar Q^\mu)\big)  F_{\mu\nu} (x+(x'-x)\tau +q)\big(\Upsilon^\nu +\tfrac{1}{\sqrt 2}(Q^\nu +\bar Q^\nu)\big)\nonumber\\
  &+im\theta\tfrac{1}{\sqrt 2}(Q^5 +\bar Q^5) \Big]~.  
\end{align}
The latter is valid for an arbitrary Abelian field $A_\mu$ and is suitable to be used also in the presence of a non trivial electromagnetic background, which can represent a macroscopic classical background and/or a set of photons. An interesting scenario is when the classical field is constant. In such a case, the worldline approach is particularly more efficient than the second-quantized counterpart, as the whole effect of constant external potential is quadratic in the worldline fields and can be fully encoded in the worldline Green's functions~\cite{Shaisultanov:1995tm}. For the scalar case, a master formula for the dressed propagator in the presence of a constant external field was derived in Ref.~\cite{Ahmad:2016vvw}.    

Here, however, we specialize to the case of a vector field which represents a number $N$ of external photons, in vacuum, i.e. in the absence of a macroscopic background. In such a case, we can rewrite~\eqref{eq:path-integral-fin} in terms of a master formula which yields the Dirac propagator with $N$ photon insertions. 

\section{$N$-photon Dirac Propagator Master Formula}
\label{sec:master-formula}

We aim to compute a generic tree level amplitudes where the fermion line is dressed with $N$ photons. The insertion of (truncated) photon lines of fixed momenta $k_l$ and polarizations $\varepsilon_l$ can be achieved within the worldline approach, by writing the vector potential as
\begin{align}
    A_\mu(x) =\sum_{l=1}^N \varepsilon_{l,\mu}\, e^{ik_l\cdot x(\tau)}~,
\end{align}
inserting it into the path integral~\eqref{eq:path-integral-fin} and singling out the multilinear part in all the $\varepsilon$'s. 
By doing this, one is led to identify the photon insertion in terms of a vertex operator, in the same way as it happens in string theory. In the present case the photon vertex operator reads (modulo the coupling constant $i{\rm e}$):
\begin{align}
    V^{x',x}[k,\varepsilon]= \int_0^1 d\tau & e^{ik\cdot (x+(x'-x)\tau +q(\tau))}
    \Bigl[  \varepsilon\cdot \big((x'-x)+\dot q(\tau)\big)\nonumber\\
    &-iT\Big(\Upsilon+\tfrac{1}{\sqrt 2}(Q(\tau)+\bar Q(\tau)) \Big) \cdot f \cdot \Big(\Upsilon+\tfrac{1}{\sqrt 2}(Q(\tau)+\bar Q(\tau)) \Big)\Bigr]~,
\end{align}
where $f_{\mu\nu}:= k_\mu \varepsilon_\nu -k_\nu \varepsilon_\mu$.  Thus, considering the insertion of $N$ photons we get the following path integral representation for the $N$-photon Dirac propagator
\begin{align}
  &S^{x',x}_N(\bar\lambda, \eta ; k_1, \varepsilon_1,\dots, k_N,\varepsilon_N)  =\sqrt 2
\int_0^\infty dT \int d\theta\, e^{-m^2 T-\tfrac{(x-x')^2}{4T}+\tfrac{1}{2T}\theta \Upsilon_\mu (x'-x)^\mu-im \theta \Upsilon^5}\nonumber\\& \int_{q(0)=0}^{q(1)=0} Dq \, e^{-\tfrac{1}{4T}\int \dot q^2}\int_{Q(0)=0}^{\bar Q(1)=0} D\bar Q DQ\, e^{-\int \bar Q \dot Q} e^{\int \big[\tfrac{1}{2\sqrt2 T}\theta ((x'-x)+\dot q(\tau))\cdot (Q(\tau)+\bar Q(\tau)) -\tfrac{i m}{\sqrt2} \theta (Q^5(\tau) +\bar Q_5(\tau))\big]}\nonumber\\
& V^{x',x}[k_1,\varepsilon_1]\cdots V^{x',x}[k_N,\varepsilon_N]~,
\end{align}
where the `dot' product represents contraction of the Greek indices and juxtaposition the contraction of capital Latin indices. The previous expression can be promptly simplified by using the  Green's function of the Grassmann coordinates, i.e.
\begin{align}
    \Big\langle Q^N(\tau) \bar Q_M(\tau')\Big\rangle = \delta^N_M \vartheta(\tau-\tau')=:\delta^N_M G_F(\tau,\tau') ~,
    \label{eq:Green-Q}
\end{align}
which inverts the kinetic operator $\partial_\tau$ and satisfies the correct boundary conditions---here $\vartheta$ is the Heaviside function. We can thus shift the Grassmann variables, in order to absorb the linear terms, namely
\begin{align}
    & Q'^\mu(\tau) = Q^\mu(\tau)+ \tfrac{\theta}{2\sqrt 2 T}  G_F\circ \dot x^\mu (\tau) =Q^\mu(\tau)+ \tfrac{\theta}{2\sqrt 2 T} \big(x^\mu(\tau)-x^\mu \big)\\
    & \bar Q'_\mu(\tau) = \bar Q_\mu(\tau)- \tfrac{\theta}{2\sqrt 2 T} \dot x_\mu \circ G_F (\tau) = \bar Q_\mu(\tau)+ \tfrac{\theta}{2\sqrt 2 T}\big(x_\mu(\tau)-x'_\mu \big)\\
    & Q'^5(\tau) =  Q^5(\tau) +G_F (\tau) \circ\tfrac{\theta m}{\sqrt 2}  \\
    & \bar Q'_5(\tau) =  \bar Q_5(\tau) -\tfrac{\theta m}{\sqrt 2}\circ G_F (\tau)
\end{align}
where the symbol `$\circ$' represents an integration over the worldline time: for example $G_F\circ \dot x^\mu (\tau) = \int d\tau' G_F(\tau,\tau') \dot x^\mu (\tau')$. By noting that the shifted fields satisfy the same boundary conditions as the original ones, and have the same kinetic operator, they thus give rise to the same Green's function as in Eq.~\eqref{eq:Green-Q}. Hence, we can just rename the new fields as the old ones, provided we perform the shift in the vertex operators, i.e.
\begin{align}
  &S^{x',x}_N(\bar\lambda, \eta ; k_1, \varepsilon_1,\dots, k_N,\varepsilon_N)  =\sqrt 2
\int_0^\infty dT \int d\theta\, e^{-m^2 T-\tfrac{(x-x')^2}{4T}+\tfrac{1}{2T}\theta \Upsilon_\mu (x'-x)^\mu-im \theta \Upsilon^5}\nonumber\\& \int_{q(0)=0}^{q(1)=0} Dq \, e^{-\tfrac{1}{4T}\int \dot q^2}\int_{Q(0)=0}^{\bar Q(1)=0} D\bar Q DQ\, e^{-\int \bar Q\cdot \dot Q} 
~ V^{x',x}[k_1,\varepsilon_1]\cdots V^{x',x}[k_N,\varepsilon_N]~,
\end{align}
where
\begin{align}
 V^{x',x}[k,\varepsilon]=& \int_0^1 d\tau  e^{ik\cdot (x+(x'-x)\tau +q(\tau))}
    e^{\varepsilon\cdot ((x'-x)+\dot q(\tau)) - iT\, \Xi(\tau,x,x')\cdot f \cdot \Xi(\tau,x,x')}\Big|_{\rm lin.\, \varepsilon}~,\\
     \Xi^\mu(\tau,x,x') :=& \Upsilon^\mu+\tfrac{1}{\sqrt 2}\big(Q^\mu(\tau)+\bar Q^\mu(\tau)\big) -\tfrac{\theta}{2T}\big(q^\mu(\tau) +(x'^\mu-x^\mu)(\tau-\tfrac12)\big)~,
\end{align}
and the path integral now only involves $Q^\mu,\ \bar Q^\mu$, and $q^\mu$ (`lin. $\varepsilon$' stands for `linear in $\varepsilon$'). The latter give rise to the bosonic Green's function
\begin{align}
  \Big \langle q^\mu(\tau)q^\nu(\tau')\Big\rangle=: \delta^{\mu\nu} G_B(\tau,\tau')=-2T\delta^{\mu\nu} \Delta(\tau,\tau'), \quad \Delta(\tau,\tau')=\tau\tau'+\frac{|\tau-\tau'|}{2}-\frac{\tau+\tau'}{2}~,
\end{align}
which satisfy (left/right bullet denotes derivative with respect to the first/second variable)
\begin{align}
    \,^\bullet \Delta(\tau,\tau')&= \tau'-\vartheta(\tau'-\tau)=\tau'+\tfrac{1}{2}\big( s_{\tau\tau'} -1\big) \nonumber \\
    \,^\bullet \Delta^\bullet(\tau,\tau')&:=1-\delta(\tau-\tau').
\end{align}
with $s_{\tau\tau'}:= {\rm sgn}(\tau-\tau')$ being the sign function. Note also that the Grassmann dynamical variables always appear in the combination $Q + \bar Q$, in terms of which
\begin{align}
    \Big\langle (Q^N(\tau)+\bar Q^N(\tau))(Q_M(\tau')+ \bar Q_M(\tau'))\Big\rangle =\delta^N_M s_{\tau\tau'} ~.
    \label{eq:Green-Q+Q}
\end{align}
Thus,
\begin{align}
  S^{x',x}_N(\bar\lambda, \eta ; k_1, \varepsilon_1,\dots, k_N,\varepsilon_N)  =&\sqrt 2
\int_0^\infty \frac{dT}{(4\pi T)^{D/2}} \int d\theta\, e^{-m^2 T-\tfrac{(x-x')^2}{4T}+\tfrac{1}{2T}\theta \Upsilon_\mu (x'-x)^\mu-im \theta \Upsilon^5}\nonumber\\& \times
 \Big\langle V^{x',x}[k_1,\varepsilon_1]\cdots V^{x',x}[k_N,\varepsilon_N]\Big \rangle\Big|_{\rm m.l.}~,
 \label{eq:S-xx'}
\end{align}
where the correlators are computed in terms of the Green's functions given above, and `m.l.' stands for `multilinear' i.e. linear in each polarization vector. 

For amplitude computations, in order to have expressions only dependent on particle
momenta, it is obviously more convenient also to Fourier transform the
end points of the fermion line, 
\begin{align}
 S^{p',p}_N(\bar\lambda, \eta ; k_1, \varepsilon_1,\dots, k_N,\varepsilon_N) = \int d^Dx d^Dx'~e^{i(p\cdot x+p'\cdot x')}  S^{x',x}(\bar\lambda, \eta ; k_1, \varepsilon_1,\dots, k_N,\varepsilon_N)~,    
\end{align}
where $p$ and $p'$ are the momenta of the incoming/outgoing Dirac fermions sitting on the fermion line; both momenta are taken to be flowing in. The previous can be performed with the convenient change of variables $x_+:=\tfrac{x+x'}2$, $x_-=x'-x$, so that the integral over $x_+$ produces a global momentum conservation delta function $(2\pi)^D \delta^{D}(p+p'+\sum_lk_l)$,  whereas the integral over $x_-$ is a Gaussian integral of the form:
\begin{align}
    \int d^D x_-~ e^{-\tfrac{1}{4T} x_-^2 +b\cdot x_-}  = (4\pi T)^{D/2}\,  e^{Tb^2}~,
\end{align}
whose numerical prefactor  cancels the denominator of Eq.~\eqref{eq:S-xx'}. Above, the vector $b^\mu$ reads
\begin{align}
    b^\mu= ip'^\mu+\sum_{l=1}^N\big(i \tau_l k_l^\mu+\varepsilon_l^\mu\big) +\frac{\theta}{2T}\Upsilon^\mu +i\theta \sum_{l=1}^N(\tau_l-\tfrac12) f_l^{\mu\nu} \big(\Upsilon_\nu+\tfrac1{\sqrt2} (Q_{l\, \nu} +\bar Q_{l\, \nu}   )\big)~.
\end{align}
Hence (the tilde sign below takes into account of the stripping off of the momentum conservation delta function),
\begin{align}
    \tilde S^{p',p}_N(\bar\lambda, \eta ; k_1, \varepsilon_1,\dots, k_N,\varepsilon_N)=\sqrt2 \int_0^\infty dT \int d\theta ~e^{-im\theta \Upsilon^5}\Big\langle e^{T(-m^2+b^2)}V[k_1,\varepsilon_1]\cdots V[k_N,\varepsilon_N]\Big\rangle\Big|_{\rm m.l.}
    \label{eq:Spp'-VN}
\end{align}
where now,
\begin{align}
    V[k,\varepsilon] =&\int_0^1d\tau\, e^{ik\cdot q(\tau) +\varepsilon\cdot \dot q(\tau) -iT\Xi(\tau)\cdot f \cdot \Xi(\tau) +i\theta q(\tau)\cdot f\cdot \Xi(\tau)}\Big|_{\rm lin.\, \varepsilon}\\
    \Xi^\mu (\tau) :=&  \Upsilon^\mu+\tfrac{1}{\sqrt 2}\big(Q^\mu(\tau)+\bar Q^\mu(\tau)\big)~.
    \label{eq:Xi}
\end{align}
In order to write down our main master formula we need to introduce some more notation which, in particular, allows us to systematically deal with the terms quadratic in the $\Xi$'s, which occur in the vertex operators. This is achieved with the help of new Grassmann numbers $\xi_l$ and $\xi'_l$, in terms of which
\begin{align}
    &\hat k^\mu_l :=\xi_l\, k^\mu_l\,,\quad \hat\varepsilon_l^\mu := \xi'_l\, \varepsilon_l^\mu\\ 
    & \tilde \varepsilon_l^\mu := \xi'_l \xi_l\, \varepsilon_l^\mu\,, \quad \tilde f^{\mu\nu}_l:=\xi'_l \xi_l\, f^{\mu\nu}_l
\end{align}
where the {\it tilded} quantities are Grassmann even, whereas the {\it hatted} ones are Grassmann odd. This can be equivalently achieved by worldline superfields, in which the above Grassmann variables are related to the supersymmetric partners of the worldline times $\tau_l$~\cite{fppaper1}.  Thus, we can write the vertex operator as
\begin{align}
 V[k,\varepsilon] =\int_0^1d\tau \int d\xi d\xi'\, e^{ik\cdot q(\tau) +\tilde\varepsilon\cdot \dot q(\tau) +i\theta q(\tau)\cdot \tilde f\cdot \Xi(\tau)+\sqrt{-i2T}(\hat k+\hat \varepsilon)\cdot \Xi(\tau) }   
\end{align}
and
\begin{align}
    \tilde S^{p',p}_N(\bar\lambda, \eta ; k_1, \varepsilon_1,\dots, k_N,\varepsilon_N)=&\sqrt2  \int_0^\infty dT \int d\theta \int d\xi d\xi'~e^{-im\theta \Upsilon^5}\nonumber\\&\times\Big\langle  V[k_1,\varepsilon_1]\cdots V[k_N,\varepsilon_N]e^{T(-m^2+\tilde b^2)}\Big\rangle
    \label{eq:Spp'-VN-xi}
\end{align}
where now $d\xi d\xi' := d\xi_1 d\xi'_1 \cdots d\xi_N d\xi'_N$ and
\begin{align}
 \tilde b^\mu= ip'^\mu+\sum_{l=1}^N\big(i \tau_l k_l^\mu+\tilde \varepsilon_l^\mu\big) +\frac{\theta}{2T}\Upsilon^\mu +i\theta \sum_{l=1}^N(\tau_l-\tfrac12) \tilde f_l^{\mu\nu}\, \Xi_{l\nu}~.  
\end{align}
so that
\begin{align}
    \tilde b^2 &= -p'^2 +\sum_{l,l'=1}^N\big( -\tau_l \tau_{l'} k_l\cdot k_{l'} +\tilde \varepsilon_l\cdot \tilde \varepsilon_{l'}+i2 \tau_l k_l\cdot \tilde\varepsilon_{l'}\big) +i2p'\cdot \sum_{l=1}^N(i\tau_l k_l +\tilde\varepsilon_l)\nonumber\\
    &+2\theta\big( ip'+\sum_{l=1}^N(i\tau_l k_l+\tilde\varepsilon_l)\big) \cdot \Big(\tfrac{1}{2T} \Upsilon+i\sum_{l'=1}^N(\tau_{l'}-\tfrac12)\tilde f_{l'}\cdot\Xi_{l'} \Big)~.
\end{align}
Most of these terms combine with the correlators of the vertex operators. In particular, let us first perform the bosonic v.e.v.'s of the product of vertex operators, i.e.
\begin{align}
   \Big\langle V[k_1,\varepsilon_1]\cdots V[k_N,\varepsilon_N]\Big\rangle_q &=e^{T\sum_{l,l'}\big(k_l\cdot k_{l'}+2\theta k_l\cdot \tilde f_{l'} \cdot \Xi_{l'} \big)\big(\tau_l\tau_{l'}+\tfrac12 \tau_{ll'} -\tfrac12 (\tau_l+\tau_{l'})\big)}\nonumber\\
   &\times e^{-T\sum_{ll'} \tilde\varepsilon_l\cdot\tilde\varepsilon_{l'}(1-d_{ll'}) }e^{-i2T\sum_{l,l'}\big(k_l+\theta \tilde f_{l}\cdot \Xi_{l}\big)\cdot \tilde \varepsilon_{l'}\big(\tau_l-\tfrac12 (s_{ll'+1})\big) }\nonumber\\
   &\times e^{\sqrt{-i2T}\sum_l(\hat k_l+\hat\varepsilon_l)\cdot \Xi_l}
\end{align}
where we have used the shortcut notations, $\tau_{ll'}:=|\tau_l-\tau_{l'}|$, $s_{ll'}:={\rm sgn}(\tau_l-\tau_{l'})$ and $d_{ll'}:=\delta(\tau_l-\tau_{l'})$. The previous combines with the $e^{T(-m^2+\tilde b^2)}$ term to give
\begin{align}
&  \Big\langle V[k_1,\varepsilon_1]\cdots V[k_N,\varepsilon_N] e^{T(-m^2+\tilde b^2)}  \Big\rangle_q =\int_0^1d\tau_1\cdots \int_0^1d\tau_N\nonumber\\
&\times e^{-T(m^2+p'^2)+T\sum_{ll'}\big(k_l\cdot k_{l'}\tfrac12 \tau_{ll'} +ik_l\cdot \tilde \varepsilon_{l'} s_{ll'}+\tilde\varepsilon_l\cdot \tilde \varepsilon_{l'}d_{ll'} \big) +iT(p'-p)\cdot \sum_l(i\tau_l k_l+\tilde \varepsilon_l)}\nonumber\\
& \times e^{\sqrt{-i2T}\sum_l(\hat k_l+\hat \varepsilon_l)\cdot \Xi_l}\, e^{T\theta \sum_l\tau_l \Xi_l\cdot \tilde f_l\cdot  (p'-p) }\, e^{-T\theta\sum_l\Xi_l\cdot \tilde f_l\cdot \big( p'+\sum_{l'}(k_{l'}\tau_{ll'}+i\tilde\varepsilon_{l'}s_{ll'})\big)  }\nonumber\\
&\times e^{\theta\big( ip'+\sum_l(i\tau_l k_l+\tilde\varepsilon_l)\big) \cdot \Upsilon}
\end{align}
where the first exponential alone would yield the scalar master formula found in Ref.~\cite{Daikouji}. Finally, we are left to perform the  fermionic v.e.v.'s, which using expressions~\eqref{eq:Xi} and~\eqref{eq:Green-Q+Q} promptly yield  
\begin{align}
  &\tilde S^{p',p}_N(\bar\lambda, \eta ; k_1, \varepsilon_1,\dots, k_N,\varepsilon_N)= \sqrt2  \int_0^\infty dT \int d\theta \int d\xi d\xi'~e^{-im\theta \Upsilon^5} \int_0^1d\tau_1\cdots \int_0^1d\tau_N\nonumber\\
&\times e^{-T(m^2+p'^2)+T\sum_{ll'}\big(k_l\cdot k_{l'}\tfrac12 \tau_{ll'} +ik_l\cdot \tilde \varepsilon_{l'} s_{ll'}+\tilde\varepsilon_l\cdot \tilde \varepsilon_{l'}d_{ll'} \big) +iT(p'-p)\cdot \sum_l(i\tau_l k_l+\tilde \varepsilon_l)}\nonumber\\
&\times e^{\sqrt{-i2T}\sum_l(\hat k_l+\hat \varepsilon_l)\cdot \Upsilon}\, e^{\tfrac{iT}{2}\sum_{ll'}s_{ll'}(\hat k_l+\hat \varepsilon_l)\cdot(\hat k_{l'} +\hat\varepsilon_{l'})}\nonumber\\
& \times  e^{\tfrac{T\theta}{2}\sqrt{-i2T}\sum_{ll'}s_{ll'}(\hat k_l+\hat\varepsilon_l)\cdot\big\{\tau_{l'}\tilde f_{l'}\cdot (p'-p)-\tilde f_{l'}\cdot \big(p' +\sum_{r'}(\tau_{l'r'}k_{r'}+is_{l'r'}\tilde \varepsilon_{r'})\big)\big\}}\nonumber\\
&\times e^{\theta\big( ip'+\sum_l(i\tau_l k_l+\tilde\varepsilon_l)\big) \cdot \Upsilon}\, e^{T\theta \sum_l\tau_l \Upsilon\cdot \tilde f_l\cdot  (p'-p) }\, e^{-T\theta\sum_l\Upsilon\cdot \tilde f_l\cdot \big( p'+\sum_{l'}(k_{l'}\tau_{ll'}+i\tilde\varepsilon_{l'}s_{ll'})\big)  }~.
\label{eq:master-N}
\end{align}
This is the final master formula for the $N$-photon Dirac propagator, where for future reference we have left the gravitino modulus $\theta$ unintegrated. On the other hand, the $\theta$-integrated version of the formula reads
\begin{align}
  &\tilde S^{p',p}_N(\bar\lambda, \eta ; k_1, \varepsilon_1,\dots, k_N,\varepsilon_N)= \sqrt2  \int_0^\infty dT  \int d\xi d\xi' \int_0^1d\tau_1\cdots \int_0^1d\tau_N\nonumber\\
&\times e^{-T(m^2+p'^2)+T\sum_{ll'}\big(k_l\cdot k_{l'}\tfrac12 \tau_{ll'} +ik_l\cdot \tilde \varepsilon_{l'} s_{ll'}+\tilde\varepsilon_l\cdot \tilde \varepsilon_{l'}d_{ll'} \big) +iT(p-p')\cdot \sum_l(i\tau_l k_l+\tilde \varepsilon_l)}\nonumber\\
&\times e^{\sqrt{-i2T}\sum_l(\hat k_l+\hat \varepsilon_l)\cdot \Upsilon}\, e^{\tfrac{iT}{2}\sum_{ll'}s_{ll'}(\hat k_l+\hat \varepsilon_l)\cdot(\hat k_{l'} +\hat\varepsilon_{l'})}\nonumber\\
& \times \Bigl\{-im\Upsilon_5+\tfrac{T}{2}\sqrt{-i2T}\sum_{ll'}s_{ll'}(\hat k_l+\hat\varepsilon_l)\cdot\Big[\tau_{l'}\tilde f_{l'}\cdot (p'-p)-\tilde f_{l'}\cdot \Big(p' +\sum_{r'}(\tau_{l'r'}k_{r'}+is_{l'r'}\tilde \varepsilon_{r'})\Big)\Big]\nonumber\\
&+ \Big( ip'+\sum_l(i\tau_l k_l+\tilde\varepsilon_l)\Big) \cdot \Upsilon + T \sum_l\tau_l \Upsilon\cdot \tilde f_l\cdot  (p'-p) -T\sum_l\Upsilon\cdot \tilde f_l\cdot \Big( p'+\sum_{l'}(k_{l'}\tau_{ll'}+i\tilde\varepsilon_{l'}s_{ll'})\Big) \Bigr\}~.
\end{align}
The previous master formula is valid off the mass-shells of photons and Dirac particles, and can thus be used as a building block to construct general spinor QED amplitudes. 

In the following section we test such formula by considering the simplest non-trivial cases.

\subsection{Special Cases: $N=0,\, 1,\, 2$}

In order to check the effective validity of the master formula proposed above, let us single out a few special cases, and map them to their corresponding expressions in terms of gamma matrices. 

Firstly, let us check that for $N=0$, i.e. no photon insertions, one recovers the free Dirac propagator. This also helps us check the overall normalization of the path integral formula. From Eq.~\eqref{eq:master-N} we simply get
\begin{align}
 \tilde S^{p',p}_0(\bar\lambda,\eta) &=\sqrt 2 \int_0^\infty dT \int d\theta ~e^{-im\theta \Upsilon^5} e^{-T(m^2+p'^2) +i\theta p'\cdot \Upsilon}   \nonumber\\
 &= i\sqrt2\,\frac{p'\cdot \Upsilon-m\Upsilon^5}{p'^2+m^2} =\frac{-i}{\langle\bar\lambda|\eta\rangle}\langle\bar\lambda|\frac{p\cdot \hat\gamma +m\hat\gamma^5} {p^2+m^2}|\eta\rangle
\end{align}
which is obviously the free Dirac propagator in the coherent-state basis. Here we have simply used the definition of the coherent states to re-instate the fermionic operators. In other words
\begin{align}
    \sqrt 2 \Upsilon^M =   \frac{1}{\langle\bar\lambda|\eta\rangle}
    \langle\bar\lambda|\hat\gamma^M|\eta\rangle~.
\end{align}
Finally, by stripping off the $\hat \gamma^5$, one is left with the operator $\hat \gamma^5 (p\cdot \hat {\tilde\gamma} +m)$, where the new gamma operators satisfy $\{ \hat{\tilde\gamma}^\mu, \hat{\tilde\gamma}^\nu\} =-2\eta^{\mu\nu}$ which is consistent with conventions adopted in Ref.~\cite{Srednicki}, in order to have a hermitian $\tilde\gamma^0$ in the presence of a mostly-positive metric signature. 

For $N=1$, we have to consider the insertion of a single photon vertex into the above path integral formula. In fact, since in this special case only equal time correlators are involved, and the equal time correlator of two $Q(\tau) + \bar Q(\tau)$ vanishes, we can drop this combination from the vertex operator, and  identify $\Xi^\mu(\tau) \to \Upsilon^\mu$. We are thus left with  
\begin{align}
 \tilde S^{p',p}_1(\bar\lambda,\eta; k,\varepsilon) & =\sqrt 2 \int_0^\infty dT \int_0^1  d\tau \int d\xi d\xi' ~ e^{-T(m^2 +p'^2+\tau (p^2 -p'^2))+iT \tilde\varepsilon \cdot (p'-p)+\sqrt{-i2T}(\hat k+\hat \varepsilon)\cdot \Upsilon}  \nonumber\\
 &\times \Big\{ -im\Upsilon^5 +i(p'+\tau k)\cdot \Upsilon +\tilde \varepsilon\cdot \Upsilon \big(1-T\tau (p^2-p'^2)\big) +Tp'\cdot k \tilde\varepsilon\cdot\Upsilon& \nonumber\\& - Tp'\cdot \tilde\varepsilon k\cdot\Upsilon +T\tau (p'-p)\cdot \tilde\varepsilon k\cdot\Upsilon \Big\}
\end{align}
where we have already integrated over the gravitino modulus, and have used the total momentum conservation. Integrating over $\xi$ and $\xi'$ which enforces the linearity in the photon polarization, and by making use of the Bianchi identity to drop a term, yields
\begin{align}
 \tilde S^{p',p}_1(\bar\lambda,\eta; k,\varepsilon) & =   \sqrt 2 \int_0^\infty dT \int_0^1  d\tau ~ e^{-T(m^2 +p'^2+\tau (p^2 -p'^2))}\Big\{T p'\cdot f \cdot \Upsilon\nonumber\\ &+T\big(\varepsilon\cdot(p'-p)-\Upsilon\cdot f \cdot \Upsilon\big) \big( m\Upsilon^5 -p'\cdot \Upsilon\big)\nonumber\\ 
 &+ \varepsilon\cdot \Upsilon \big(1-T\tau(p^2-p'^2) \big)\Big\}~.
\end{align}
The integrands in the first two lines are $\tau$-independent; thus, for all of them, the time integrals just produce the denominator $(p'^2+m^2)(p^2+m^2)$. The last term involves terms of the form
\begin{align}
    \int_0^1  d\tau ~ e^{a \tau} (1+a\tau) =(1+a\partial_a) \int_0^1  d\tau ~ e^{a \tau} =e^a\,,\quad a=-T (p^2-p'^2)~,
    \label{eq:cancel}
\end{align}
which, integrated over the Schwinger time $T$, simply gives $\tfrac1{p^2+m^2}$. However, note that the two terms in the integrand of Eq.~\eqref{eq:cancel} would separately lead to contributions which behave as $1/T$, and which would diverge in the $T$ integral. However, these would-be singular terms cancel each other in the combination~\eqref{eq:cancel}. Thus, the final expression reads
\begin{align}
 \tilde S^{p',p}_1(\bar\lambda,\eta; k,\varepsilon) =   \frac{\sqrt 2}{(p'^2+m^2)(p^2+m^2)}\Big\{   & p'\cdot f \cdot \Upsilon +\big(\varepsilon\cdot(p'-p)-\Upsilon\cdot f \cdot \Upsilon\big) \big( m\Upsilon^5 -p'\cdot \Upsilon\big)\nonumber\\&+ \varepsilon\cdot\Upsilon (p'^2+m^2)\Big\}~.
 \label{eq:Spp'-fin}
\end{align}
In order to complete our test, we relate the previous to the Dirac propagator written in the spin (Fock) basis, i.e. we  map the products of $\Upsilon$'s to (products of) gamma matrices. Above we have written the map for a single $\Upsilon$. However, it is easy to show that a product of an arbitrary number of $\Upsilon$'s, due to their Grassmannian nature, corresponds to the antisymmetric product of gamma matrices, i.e.
\begin{align}
    \Upsilon^{M_1}\cdots \Upsilon^{M_k} =\tfrac{1}{2^{k/2}}\frac{\langle \bar \lambda | \hat \gamma^{M_1\cdots M_k}|\eta\rangle}{\langle \bar \lambda| \eta\rangle} \ \longrightarrow \ \frac{1}{2^{k/2}} \gamma^{M_1\cdots M_k}
    \label{eq:identify}
\end{align}
where the juxtaposition of indices in the gamma matrix product means antisymmetrization. In the last step, in order to simplify the notation, we have stripped off the coherent state basis vectors and omitted the operator symbol in the product of gamma matrices. In particular, the products that appear in Eq.~\eqref{eq:Spp'-fin} are
\begin{align}
    &\Upsilon^M \Upsilon^N \ \longrightarrow\ \tfrac{1}{2}\gamma^{MN} = \tfrac12 \big( \gamma^M \gamma^N -\eta^{MN}\big)\\
    &\Upsilon^M \Upsilon^N \Upsilon^R \ \longrightarrow\ \tfrac{1}{2^{3/2}}\gamma^{MNR}=\tfrac{1}{2^{3/2}}\big(\gamma^M \gamma^N\gamma^R -\eta^{MN}\gamma^R +\eta^{MR}\gamma^N-\eta^{NR}\gamma^M \big)~.
\end{align}
Thus, using such an identification in the expression~\eqref{eq:Spp'-fin}, and using the Clifford algebra, we obtain
\begin{align}
     \tilde S^{p',p}_1(\bar\lambda,\eta; k,\varepsilon) \ \longrightarrow \ (-i)^2\frac{(-\slashed{p'}+m\gamma^5)\slashed{\varepsilon}(\slashed{p}+m\gamma^5)}{(p'^2+m^2)(p^2+m^2)}~,
\end{align}
which is the Dirac propagator for an incoming electron of momentum $p$, an outgoing electron of momentum $-p'$ flowing out, and the insertion of a photon of momentum $k=-(p+p')$ and polarization~$\varepsilon$.~\footnote{Note that, throughout the paper, the `slash' denotes a contraction with respect to the $\gamma^\mu$, not with respect to the $\tilde\gamma^\mu$.}

Note that, in the previous computation, we have never used the photon on-shell condition and no gauge choice has been made. In fact, as already stressed, our master formula is valid off-shell.

Finally, let us check the validity of the master formula~\eqref{eq:master-N} for $N=2$, i.e. for the tree level Compton scattering. Here we will content ourselves and impose on-shell condition and Lorentz gauge to the photons, $k^2=k\cdot \varepsilon =0$. We thus obtain
\begin{align}
    &\int_0^\infty dT \int d\theta \int d\tau_1 d\tau_2 \, e^{-T(p'^2+\bp \cdot (k_1 \tau_1+k_2 \tau_2)+m^2)+i\theta\bigr((p'+k_1 \tau_1+k_2 \tau_2)\cdot \Upsilon-m\psi^5\bigr) +Tk_1 \cdot k_2\tau_{12}} \Bigr\{ \nonumber \\
    &-2T^2 \Upsilon \cdot f_1 \cdot f_2 \cdot \Upsilon s_{12}-\frac{T^2}{2}\mathrm{tr}(f_1 f_2)+2T\varepsilon_1 \cdot \varepsilon_2 \delta(\tau_1-\tau_2)-iT\varepsilon_1 \theta \cdot f_2 \cdot \Upsilon s_{12}-iT \varepsilon_2 \theta \cdot f_1 \cdot \Upsilon s_{21} \nonumber \\
    &+\bigr(\varepsilon_1 \theta \cdot \Upsilon +iT \varepsilon_1 \cdot \bp -iT\varepsilon_1 \cdot k_2 s_{12}-iT \Upsilon \cdot f_1 \cdot \Upsilon\bigr) \bigr(1 \leftrightarrow 2\bigr) \nonumber \\
    &-i\theta T^2 \bigr(\bp \cdot \varepsilon_1 - \varepsilon_1 \cdot k_2 s_{12}- \Upsilon \cdot f_1 \cdot \Upsilon \bigr)\Upsilon \cdot f_2 \cdot \bigr(p'+k_1\tau_{12}-\bp\tau_2 \bigr) \nonumber \\
    &-i\theta T^2 \bigr(\bp \cdot \varepsilon_2 - \varepsilon_2 \cdot k_1 s_{21}- \Upsilon \cdot f_2 \cdot \Upsilon \bigr)\Upsilon \cdot f_1 \cdot \bigr(p'+k_2 \tau_{12}-\bp\tau_1 \bigr) \nonumber \\
    &+i\theta T^2\Upsilon \cdot f_1 \cdot f_2 \cdot \bigr(p'+k_1\tau_{12}-\bp\tau_2 \bigr)s_{12}+i\theta T^2 \bigr(p'+k_1\tau_{12} -\bp\tau_1 \bigr) \cdot f_1 \cdot f_2 \cdot \Upsilon s_{12} \Bigr\}~,
    \label{eq:formula-N=2}
\end{align}
where we have used the $\bp:=p'-p$.

In expression~\eqref{eq:formula-N=2}, similarly to the $N=1$ case, one can see that some of the terms are individually divergent upon the integral over the Schwinger time. However, these potentially divergent contributions cancel each other already at the integrand level, analogously to what happens in Eq.~\eqref{eq:cancel}~\cite{DegliEsposti-LM-thesis}.

After all the Grassmann integrals are computed, Eq.~\eqref{eq:formula-N=2} becomes
\begin{align}
    &\int_0^\infty dT \int d\tau_1 d\tau_2 \, e^{-T(p'^2+\bp \cdot (k_1 \tau_1+k_2 \tau_2)+m^2) +Tk_1 \cdot k_2\tau_{12}}\Bigl\{ \nonumber \\
    &\Bigr(-2T^2 \Upsilon \cdot f_1 \cdot f_2 \cdot \Upsilon s_{12}-\frac{T^2}{2}\mathrm{tr}(f_1 f_2)+2T\varepsilon_1 \cdot \varepsilon_2 \delta(\tau_1-\tau_2) \Bigr)i\bigr( p'\cdot \Upsilon-m\Upsilon^5 \bigr) \nonumber \\
    &-iT^2(\bp \cdot \varepsilon_1 -\varepsilon_1 \cdot k_2 s_{12}-\Upsilon \cdot f_1 \cdot \Upsilon)(1 \leftrightarrow 2)\bigr( p' \cdot \Upsilon-m\Upsilon^5 \bigr) \nonumber \\
    &+iT^2 \bigr(\bp \cdot \varepsilon_1 - \varepsilon_1 \cdot k_2 s_{12}- \Upsilon \cdot f_1 \cdot \Upsilon \bigr)p' \cdot f_2 \cdot \Upsilon +iT^2\Upsilon \cdot f_1 \cdot f_2 \cdot p' \, s_{12} \Bigr\} \nonumber \\
    &+iT\int_0^\infty dT e^{-T(p'^2+m^2+\bp \cdot k_1 -k_1 \cdot k_2)} \int_0^1 d\tau_2 \, e^{-T(\bp \cdot k_2 +k_1 \cdot k_2)\tau_2} \Bigr(\varepsilon_1 \cdot \Upsilon (-2p \cdot \varepsilon_2 -\Upsilon \cdot f_2 \cdot \Upsilon) \nonumber \\
    &-\varepsilon_1 \cdot f_2 \cdot \Upsilon \Bigr) \nonumber \\
    &+iT\int_0^\infty dT e^{-T(p'^2+m^2+\bp \cdot k_2 -k_1 \cdot k_2)} \int_0^1 d\tau_1 \, e^{-T(\bp \cdot k_1 +k_1 \cdot k_2)\tau_1} \Bigr(\varepsilon_2 \cdot \Upsilon (2p' \cdot \varepsilon_1 -\Upsilon \cdot f_1 \cdot \Upsilon) \nonumber \\
    &-\varepsilon_2 \cdot f_1 \cdot \Upsilon \Bigr).
\end{align}
For the sake of comparison, it is helpful to identify the contributions of single Feynman diagrams to the amplitude which,
as usual in the worldline approach, is a straightforward task. For example, considering the following contribution associated to the time ordering $\tau_1>\tau_2$, we get 
\begin{align}
    &\int_0^\infty dT \, T^2 e^{-T(p'^2+m^2)} \int_0^1 d\tau_1 \int_0^1 d\tau_2 \, e^{-T\bp \cdot (k_1 \tau_1+k_2 \tau_2)+Tk_1 \cdot k_2 |\tau_{12}|} \vartheta(\tau_1-\tau_2)= \nonumber \\
    &\frac{1}{(p'^2+m^2)(p^2+m^2)[(p'+k_1)^2+m^2]} =\frac{1}{(p'^2+m^2)(p^2+m^2)[(p+k_2)^2+m^2]}
    \end{align}
    which is the correct denominator for one of the two Feynman diagrams of the spinor tree-level Compton scattering. Similarly, taking the opposite time ordering one gets a similar expression with $k_1$ replaced by $k_2$, and viceversa. Therefore, for the sake of extracting single Feynman diagrams, it is sufficient to express `sign' functions in terms of Heaviside's. Moreover, the terms which have single $\tau$ integrals give denominators with double poles, one with $(p'^2+m^2)((p+k_1)^2+m^2)$, the other with $k_1\to k_2$.
Thus, let us collect all terms with momentum $p+k_2$ in the denominator, i.e.
\begin{align}
    &\frac{i}{(p'^2+m^2)(p^2+m^2)[(p+k_2)^2+m^2]}\Bigl\{ \Bigl(-2\Upsilon \cdot f_1 \cdot f_2 \cdot \Upsilon -\frac{1}{2}\mathrm{tr}(f_1 f_2) \Bigr)\bigr( p'\cdot \Upsilon-m\Upsilon^5 \bigr) \nonumber \\
    &-(\bp \cdot \varepsilon_1 -\varepsilon_1 \cdot k_2 - \Upsilon \cdot f_1 \cdot \Upsilon)(\bp \cdot \varepsilon_2 +\varepsilon_2 \cdot k_1 -\Upsilon \cdot f_2 \cdot \Upsilon)\bigr( p' \cdot \Upsilon-m\Upsilon^5 \bigr) \nonumber \\
    &+\bigr(\bp \cdot \varepsilon_1 - \varepsilon_1 \cdot k_2 -\Upsilon \cdot f_1 \cdot \Upsilon \bigr)p' \cdot f_2 \cdot \Upsilon + \bigr(\bp \cdot \varepsilon_2 + \varepsilon_2 \cdot k_1- \Upsilon \cdot f_2 \cdot \Upsilon \bigr)p' \cdot f_1 \cdot \Upsilon \nonumber \\
    &+\Upsilon \cdot f_1 \cdot f_2 \cdot p' \, - p' \cdot f_1 \cdot f_2 \cdot \Upsilon +\varepsilon_1 \cdot \Upsilon (\bp \cdot \varepsilon_2 +\varepsilon_2 \cdot k_1 -\Upsilon \cdot f_2 \cdot \Upsilon)(p'^2+m^2) \nonumber \\
    &-\varepsilon_1 \cdot f_2 \cdot \Upsilon(p'^2+m^2) \Bigr\}~,
\end{align}
which, using the map~\eqref{eq:identify}, yields 
\begin{align}
    &\frac{i(\ssl p' - m \gamma^5)\ssl \varepsilon_1 (-\ssl p - \ssl k_2 - m\gamma^5) \ssl \varepsilon_2 (\ssl p+m \gamma^5)}{(p'^2+m^2)(p^2+m^2)[(p+k_2)^2+m^2]}-\frac{i(\ssl p' - m \gamma^5)\ssl \varepsilon_1 \ssl \varepsilon_2}{(p'^2+m^2) (p^2+m^2)}~,
    \label{eq:pk1}
\end{align}
whereas the `symmetric' contributions associated to the replacement $1 \leftrightarrow 2$, are
\begin{align}
    &\frac{i(\ssl p' - m \gamma^5)\ssl \varepsilon_2 (-\ssl p - \ssl k_1 - m\gamma^5) \ssl \varepsilon_1 (\ssl p+m \gamma^5)}{(p'^2+m^2)(p^2+m^2)[(p+k_1)^2+m^2]}-\frac{i(\ssl p' - m \gamma^5)\ssl \varepsilon_2 \ssl \varepsilon_1}{(p'^2+m^2) (p^2+m^2)}~.
    \label{eq:pk2}
\end{align}
Finally, the term with the delta function  gives
\begin{align}
    \frac{2i(\ssl p' - m\gamma^5)\varepsilon_1 \cdot \varepsilon_2} {(p'^2+m^2)(p^2+m^2)}
\end{align}
which precisely cancels the last two terms in equations~(\ref{eq:pk1}, \ref{eq:pk2}). Overall, we finally obtain
\begin{align}
    (-i)^3(iq)^2\frac{-\ssl p' + m \gamma^5}{p'^2+m^2} \biggr( \ssl \varepsilon_1 \frac{\ssl p + \ssl k_2 + m \gamma^5}{(p+k_2)^2+m^2}\ssl \varepsilon_2 + \ssl \varepsilon_2 \frac{\ssl p + \ssl k_1 + m \gamma^5}{(p+k_1)^2+m^2} \ssl \varepsilon_1 \biggr) \frac{\ssl p + m \gamma^5}{p^2+m^2}
\end{align}
which is exactly the (untruncated) Compton scattering amplitude.


\section{Conclusions and Outlook}
In the present manuscript we have introduced a worldline first-principled way to compute the Dirac propagator dressed with an arbitrary number of external photons, making use of a locally supersymmetric ${\cal N}=1$ spinning particle model, which allows to represent the numerator of the Dirac propagator directly from the worldline action, and representing the spinorial degrees of freedom by means of a coherent state basis. The main result consists in a compact master formula akin to those derived earlier for one-loop scattering amplitudes, and more recently for tree-level scattering amplitudes, with a scalar line and a fermion line. Such master formulas, at each $N$, provide the full scattering amplitude, i.e. the sum of the different Feynman diagrams involved in a given process. At loop level this property was exploited in the computation of higher loop contributions to the Euler-Heisenberg lagrangians~\cite{Reuter:1996zm, Fliegner:1997ra}. The tree level master formula may instead turn out to be helpful for higher loop $g-2$ computations.  

There are several natural extensions to the present computation, which we leave for future work. Firstly, it would be helpful to generalize the master formula including the coupling to a non-Abelian external field~\cite{Ahmadiniaz:2015xoa} and/or to gravity: to the best of our knowledge, no such derivations exist in the literature. Another intriguing avenue would be  to pack together the coherent state approach to the spinor helicity formalism, which has proven to be quite efficient for on-shell    scattering amplitude computations. Finally, it would be a welcome addition to extend  the present approach to the dressed Dirac propagator in the presence of an external background field, for example a constant electromagnetic field, such as it was done in Ref.~\cite{Ahmad:2016vvw} for the scalar line. Such extension should be relatively straightforward, yet computationally trickier, since the Green's functions become matrix-valued.     

\paragraph{Acknowledgements}
The Authors thank Naser Ahmadiniaz, Fiorenzo Bastianelli and Christian Schubert for helpful comments and suggestions.


\end{document}